\documentclass[aps,prd,twocolumn,groupedaddress,showpacs,preprintnumbers,draft]
{revtex4}
\usepackage{graphicx}
\usepackage{epsf} 

\newcommand{\ba}{\begin{array}}
\newcommand{\ea}{\end{array}}
\def\be{\begin{equation}}
\def\ee{\end{equation}}
\def\bea{\begin{eqnarray}}
\def\eea{\end{eqnarray}}
\def\bs{\begin{subequations}}
\def\es{\end{subequations}}

\def\l{\lambda}

\def\d{\partial}

\def\a{\alpha}

\newcommand{\bad}{\begin{array}{ccc}}

\newcommand{\ra}{\rightarrow}
\newcommand{\vs}{\vspace{5mm} \noindent}

\begin{document}

\preprint{MCGILL-04-05} 
\preprint{UdeM-GPP-TH-05-134}


\title{Dynamical CP Violation in the Early Universe and Leptogenesis} 

\author{K.R.S. Balaji $^{1)}$ \email[Email:]{balaji@hep.physics.mcgill.ca}
Tirthabir Biswas $^{1)}$ \email[Email:]{tirtho@hep.physics.mcgill.ca},
Robert H. Brandenberger $^{1,2)}$ \email[Email:]{rhb@het.brown.edu}
and David London $^{3)}$ \email[Email:]{london@lps.umontreal.ca}}
\affiliation{1) Department of Physics, McGill University, Montr\'eal, QC, 
H3A 2T8, CANADA}
\affiliation{2) Department of Physics, Brown University, Providence, RI 02912, 
USA} 
\affiliation{3) Laboratoire Ren\'e J.-A. L\'evesque, 
Universit\'e de Montr\'eal, C.P. 6128, succ. centre-ville, Montr\'eal, QC,
H3C 3J7, CANADA.}
 
\begin{abstract}
In a recent publication, we suggested a mechanism for obtaining
dynamical CP violation in the early Universe based on the
out-of-equilibrium evolution of complex scalar fields. In this paper,
we suggest several ways of transferring the CP asymmetry from the
scalar sector to the leptonic sector. In particular, we point out how
a ``transient MNS(P) matrix'' can generate an asymmetry between
fermions and anti-fermions directly.
\end{abstract}

\maketitle

\section{Introduction}

CP violation is one of the key ingredients required for baryogenesis
\cite{Sakharov}. In most particle-physics models, CP violation is
explicit in the Lagrangian. In the Standard Model (SM), CP violation
is built in via complex phases in the Cabibbo-Kobayashi-Maskawa (CKM)
matrix \cite{CKM}, albeit with a very small amplitude. Since SM CP
violation is very small, it is not possible to obtain a sufficiently
large net baryon to entropy ratio in this scenario. Hence, in
extensions of the SM additional CP violation is often introduced via
CP-violating phases in an extended Higgs sector to explain
lepto/baryogenesis.

Following earlier ideas of Dolgov \cite{dolgov}, we recently proposed
an alternative dynamical mechanism of CP violation in the early
Universe \cite{bala1} (see also \cite{bala2}).  Here we elaborate on
this mechanism. In this scenario, the Lagrangian is CP
symmetric. However, there are complex phases which arise as initial
conditions for scalar fields in the early Universe. In the framework
of inflationary cosmology it is possible that these phases are
coherent over the present Hubble patch, as discussed in
\cite{bala1}. These complex phases lead to CP violation. The magnitude
of CP violation decreases as the scalar fields relax to their ground
state. Thus, in the framework of dynamical cosmological CP violation,
large CP violation in the early Universe leading to the observed
baryon to entropy ratio is no longer in conflict with the small
magnitude of the presently measured CP violation in the laboratory.

In \cite{bala1}, we discussed the general framework of dynamical
cosmological CP violation in the case of a model with two complex
scalar fields. We showed explicitly how the dynamics of the background
scalar fields (the ``condensates'') can lead to the generation of a CP
asymmetry in the scalar field quanta which are generated via the decay
of the condensates. We briefly mentioned how this CP asymmetry in the
scalar sector can be transformed into a CP asymmetry of the fermions
leading eventually to baryogenesis \footnote{Note that in terms of
using dynamical scalar fields, our mechanism has similarities with the
Affleck-Dine scenario \cite{AD}. The novel aspect is that we do not
introduce explicitly CP and baryon number violating terms in the
Lagrangian.}. In this paper we elaborate on these two issues. Moreover
we point out a second channel to produce CP violation in the fermionic
sector directly from the dynamics of background scalar fields. We
observe that non-trivial phases of the evolution of the condensate can
induce complex Yukawa terms in the Lagrangian. This results in what we
call the `transient Maki-Nakagawa-Sakata (Pontecorvo) MNS(P) matrix''
\cite{MNSP}, and leads to CP violation in the leptonic sector.

The outline of this paper is as follows: in Section 2 we review the
basic ideas of dynamical cosmological CP violation as discussed in
\cite{bala1} and give a brief overview of the ways in which this can
result in lepto/baryogenesis. In Section 3 we study the avenues for
leptogenesis which make use of an asymmetry in scalar field quanta
generated by the evolving background condensate. In Section 4 we study
leptogenesis channels in which the background condensate directly
induces the lepton asymmetry via the transient MNS(P) matrix. In
Section 5, we consider a specific leptogenesis channel and evaluate
how the resulting lepton asymmetry depends on the parameters and
initial conditions of our model. In section 6, we comment on other
possibilities of realizing leptogenesis using complex initial
conditions. We conclude in Section 7 with a summary of the main
results and a discussion of future directions of research.

\section{Dynamical CP Violation}

In this section, we review the scenario of \cite{bala1} in which
nontrivial phases in the initial conditions of scalar fields lead to
dynamical CP violation in a theory in which the Lagrangian is
symmetric under CP. We then explain how this dynamical CP violation
can be transferred to fermions, leading to a lepton-anti-lepton
asymmetry, and eventually to a baryon-antibaryon asymmetry.

\vs {\bf Phases from Initial Conditions:} Consider a toy model
containing the two complex doublet scalar fields $\phi_1$ and $\phi_2$
(we take doublets instead of singlets only to make the field content
of the model look more like that of the Standard Model). For the
moment, we say nothing about the quantum numbers of these two fields,
except that they are the same for both $\phi_i$, $i=1,2$. The scalar
potential of the model is taken to be
\be
V(\phi_1,\phi_2) = \sum_{i=1,2} m_i^2 \phi_i^\dagger \phi_i +
V_4(\phi_1, \phi_2)~,
\label{pot}
\ee
where it is assumed that $m_1 > 2 m_2$, and
\be
V_4(\phi_1,\phi_2) = g (\phi_2^\dagger\phi_1) (\phi_2^\dagger \phi_2) +
h.c. ~,
\label{pot2}
\ee
where $g$ is a real coupling constant.

It is natural to assume that only the neutral components of $\phi_i$
pick up an initial non-vanishing expectation value \footnote{If an
electrically-charged scalar obtains an initial condition, we break
electromagnetism, in a way similar to spontaneous symmetry
breaking. Given the neutrality of our Universe, and strong constraints
on the isotropy of the cosmic microwave background radiation
\cite{caprini}, we presume that electro-magnetism was never broken
during the course of early universe.}. From now on the $\phi_i$'s will
denote the neutral components of the respective doublets. Note that
their potential will look exactly the same as Eqs.~(\ref{pot}) and
(\ref{pot2}), with the hermitian conjugates replaced by complex
conjugates of the neutral components.
 
In the context of hot big bang or inflationary cosmology, it is
natural to assume that the neutral scalar fields will start out
displaced from their vacuum values. In a non-inflationary Universe,
the spatial gradients of these fields would be large and they would
rapidly relax to their ground-state values. However, in the context of
inflationary cosmology the situation is very different. If the mass of
the neutral scalar fields is smaller than the Hubble constant during
inflation, the fields will evolve essentially as free scalar
fields. That is, they will respond to quantum fluctuations like a free
scalar field, and thus acquire a large root mean square expectation
value when averaged over a region of radius $H^{-1}$ (a ``Hubble
patch''). At the end of inflation, these neutral scalar fields will
thus in general be displaced from the minimum of the bare potential
energy function by an amount which is large compared to $H$ (see
e.g.~\cite{Linde,FordVil} and \cite{Lindebook} for a review).

After inflation ends, the Hubble constant will gradually decrease,
eventually falling below the mass scale of the scalar fields. At this
point, these scalar fields will begin to roll towards their ground
state. It is this dynamics in the post-inflationary Universe which
will lead to dynamical CP violation and resulting leptogenesis. To see
this, observe that although one of the two phases of the two neutral
complex scalar fields can in general be eliminated by a phase 
redefinition, the second
cannot. If we choose the initial value of $\phi_2$ to be real (and
denoted by $a_2$), but take the initial value of $\phi_1$ to have a
phase $\alpha$, i.e.\ to be $a_1 e^{i \alpha}$, then in order to
analyze the particle-physics processes one needs to consider
fluctuations of the scalar fields around this complex initial
condition (background). This effectively introduces non-trivial
complex phases in the scalar potential as well as in the Yukawa
interactions with the fermions, leading to CP violation. In this paper
we examine ways in which such CP violation can be transferred to the
fermion sector, concentrating principally on leptogenesis scenarios

\vs {\bf Baryogenesis preceded by Leptogenesis:} In order to produce a
baryon-antibaryon asymmetry we first of all need processes which
violate baryon number. Secondly, we require C and CP violation
\cite{Sakharov} (For recent reviews on baryogenesis, we refer the
reader to \cite{breview}). In most baryogenesis scenarios, the
required CP violation is explicitly introduced into the Lagrangian.
However, as discussed earlier, in our approach this is not the
case. Rather, it is the initial conditions which generate this
asymmetry. This dynamical mechanism of CP violation can be coupled to
many of the well-known models of baryogenesis. However, in light of
current data which points to a nonzero neutrino mass \cite{nudata},
one of the preferred mechanisms for generating the baryon asymmetry
involves first generating a lepton asymmetry. This asymmetry then can
be converted to a baryon-antibaryon asymmetry via
baryon-number-violating sphaleron processes \cite{sphaleron}. As we
emphasize below, this mechanism does not require any new baryon or
lepton-number-violating terms in the Lagrangian.  In the following, we
shall focus on this possibility. (Note that, in this paper, we do not
discuss the possibility of baryogenesis via quarks alone. However, it
is straightforward to modify our scenarios to include CP violation in
the quark sector directly.)

Note that it is not necessary to have more leptons than anti-leptons
in order for the sphalerons to produce more baryons than anti-baryons.
In fact, an asymmetry in only certain states is sufficient. This
observation comes about because the weak interactions -- and hence the
sphaleron interactions -- couple to only left-handed particles and
right-handed anti-particles. Since only these states are involved, the
production of $\nu_R$ or ${\bar\nu}_L$ cannot lead to a
baryon-antibaryon asymmetry. While it is true that a $\psi_L$ state
can be converted into a $\psi_R$ state, the rate is proportional to
the mass of the $\psi$. For light neutrinos, this conversion rate is
sufficiently slow that an asymmetry between $\nu_L$ and ${\bar\nu}_R$
survives to the weak scale. Thus, sphalerons can convert an asymmetry
between $\nu_L$ and ${\bar\nu}_R$ alone into a baryon asymmetry.

On the other hand, this does not hold for electrons since any
generated asymmetry is washed out by the mass terms \cite{bcamp,cline}.
Thus, an asymmetry between $e_L$ and
${\bar e}_R$ cannot be converted into a baryon-antibaryon asymmetry
via the standard electroweak sphaleron effects.

The upshot of this discussion is that we can successfully obtain
baryogenesis involving sphalerons provided it is preceded by one of
two scenarios. Either (i) we have full leptogenesis, i.e.\ there is an
asymmetry between lepton and anti-lepton number, or (ii) an asymmetry
between $\nu_L$ and ${\bar\nu}_R$ is created. The former case
naturally suggests that there are more leptons than anti-leptons.. In
this case of course one requires the presence of
lepton-number-violating processes, as is required in thermal
leptogenesis scenarios \cite{fuku}, for example. The latter scenario
is in spirit similar to Dirac leptogenesis \cite{manfred}, but it
differs in two aspects. First, there need not be explicit
lepton-number violation, and second, apriori the Lagrangian is CP
conserving.

In the following sections we discuss two distinct mechanisms for
generating a lepton asymmetry. The first involves CP-violating scalar
interactions which generate an asymmetry between scalar particles and
their anti-particles. This CP asymmetry is then transferred to the
lepton sector. The second uses what we call the ``transient MNS(P)
matrix.'' Note that, for either mechanism to work, one needs Yukawa
interactions between the rolling scalar fields and the SM fermions.
However, the specific applications of these interactions depend on the
mechanism, and are slightly different.

\section{Lepton Asymmetry from an Asymmetry of Scalar Quanta}

\noindent
{\bf CP Violation in the Scalar Sector:} We begin by considering
mechanisms by which a CP asymmetry involving scalars can be
transferred to the lepton sector. As discussed in \cite{bala1} (and
making use of the phase conventions discussed in Section 2) we
consider field fluctuations about the initial conditions:
\be
\phi'_2 = \phi_2 - a_2 ~~,~~~~ \phi'_1 = \phi_1 - a_1 e^{i\alpha} ~.
\label{initc}
\ee
Note that in the context of cosmologically evolving fields, the
background fields $a_{i}$ vary with time. We shall address the
dynamical evolution later (in Section 5). For now, we mention that, in
order for the CP-violating effects at different times to add up
coherently, we work in the adiabatic approximation where we assume
that the time-dependence of the background fields is slow compared to
the interaction time scale of the fluctuations. We will focus on the
time-dependence of the amplitude of the fields. However, in principle,
the phase $\alpha$ can also be time dependent \cite{raghu}.

The relations (\ref{initc}) can be inverted and inserted into the
scalar potential of (\ref{pot}). Expanding the quartic term of
(\ref{pot2}), one finds both quadratic and cubic terms. Dropping the
primes, the cubic interaction terms become
\be
V_3 = g \left[ a_1 e^{i\alpha} \phi_2^\dagger |\phi_2|^2 + a_2
\phi_2^\dagger \phi_2^\dagger \phi_1 + 2 a_2 |\phi_2|^2 \phi_1 \right]
+ h.c.
\label{cubic2}
\ee
The induced couplings contribute to the decay $\phi_1 \to \phi_2
\phi_2$ \cite{bala1}. This decay is CP-violating, as can be seen as
follows. There are two types of diagrams: a tree diagram and several
loop diagrams. Because these diagrams have relative weak and strong
phases, one finds that the number of $\phi_2$ states is unequal to the
number of $\phi^\dagger_2$ states \footnote{We have implicitly assumed
that the quadratic corrections from the initial conditions do not
change the mass eigenstates appreciably. This assumption however is
not necessary. In general one has to compute the asymmetry in the
current mass basis and then rotate it back to the original
$\phi_1,\phi_2$ basis. This only leads to technical complications but
does not change the basic physical effect.}. This is CP violation in
the scalar sector. Specifically, the relative asymmetry $A_{CP}$ in
the decay rates is
\be
A_{CP} \sim \sin 2\alpha ~.
\label{sacp}
\ee

This asymmetry in the scalar sector must now be transferred to the
leptonic sector. This happens through the Yukawa interactions. As
discussed, earlier, there are two possibilities. Either we have full
leptogenesis, or an asymmetry between $\nu_L$ and ${\bar\nu}_R$ alone
is created. We refer to these as scenarios A and B, respectively.

\vs {\bf Transferring Asymmetry to Fermions:} We begin with mechanism
A and briefly discuss the generation of a CP asymmetry in the leptonic
sector via the decay of scalar quanta. In this case one requires
lepton-number-violating processes in which the asymmetry in the CP
values of the scalar excitations, computed above, transfers directly
into an equivalent asymmetry in the leptonic sector. In the case of
Dirac neutrinos, the generic CP-violating rate asymmetry is of the
form
\bea
\Delta \Gamma_{\phi_i \to f_1f_2...}  &=& \Gamma_{\phi_i \to
f_1f_2...} - \Gamma_{\phi_i^* \to \bar f_1 \bar f_2...}  \nonumber \\
&\simeq& \Gamma_{\phi_i \to f_1f_2...}A_{CP}~,
\label{dacp}
\eea
where $\Gamma_{\phi_i \to f_i ...}$ is the decay rate of a scalar
$\phi_i$ to a final state involving Dirac fermions $f_i$. Here,
$A_{CP}$ denotes the inherent CP asymmetry in the production of
scalars $\phi_i$ and its CP conjugated state, $\phi_i^*$. Note that an
asymmetry cannot be generated for Majorana neutrinos. This is because
in this case the final states are not distinguishable as particles and
anti-particles, which is required to obtain an asymmetry.

We now consider an example of scenario B, in which one generates only
an asymmetry between $\nu_L$ and ${\bar\nu}_R$. Consider the following
Yukawa interactions between neutrinos and the doublets $\phi_i$:
\be
{\cal L}_y = \frac12 d_{\nu i} {\bar\nu} (1 + \gamma_5) \nu \phi_i +
h.c. ~.
\label{LDirac}
\ee
This leads to the decays (we consider only diagonal couplings here)
\be
\phi_2 \to {\bar\nu}_L \nu_L ~~,~~~~ \phi_2^\dagger \to {\bar\nu}_R
\nu_R ~.
\label{sdecay}
\ee
The asymmetry in the number of $\phi_2$ and $\phi^\dagger_2$ fields
will therefore translate into an asymmetry in the number of
${\bar\nu}_L \nu_L$ and ${\bar\nu}_R \nu_R$ final states. As explained
above, since only the states $\nu_L$ and ${\bar\nu}_R$ interact via
the weak interactions, sphalerons will convert these states into
baryons. In other words, this will give rise to a baryon asymmetry.

This asymmetry is easily estimated to be
\be
Y_B = \frac{22}{79} Y_\nu ~,
\label{basym}
\ee
where we have assumed that only the standard model degrees of freedom
are involved in the sphaleron process \cite{turner}. The
neutrino/lepton asymmetry $Y_\nu$ is estimated from the asymmetry in
the production rate of the $\phi_2$ field. Following (\ref{sacp}) and
(\ref{dacp}) we have the lepton asymmetry
\be
Y_\nu \propto \sin\alpha \, \Gamma_{\phi_2 \to {\bar\nu}_L \nu_L } 
\sim \frac{1}{8 \pi} \sin \alpha \, d_{\nu 2}^2 \, m_2 ~,
\ee
so that
\be
Y_B \sim \frac{22}{79}\frac{1}{8 \pi} \sin \alpha \, d_{\nu 2}^2 \, m_2~.
\label{dasym}
\ee

The requirement that this asymmetry not be washed out is given by the
condition \cite{sfb}
\be
\frac{m_2^2 \, m_\nu^2 \, \Gamma_{\phi_2 \to {\bar\nu}_L \nu_L}}{2T^4
\, H(m_2) } \ll 1~.
\label{cond}
\ee
In (\ref{cond}), $T$ denotes the temperature and $H(m_2)$ is the value
of the Hubble constant at scale $m_2$, when the scalar decays to
neutrinos of mass $m_\nu$. Note that this condition is obtained by
solving for the relevant Boltzmann equations (for details, we refer the
reader to \cite{sfb}) and is easily satisfied because of the smallness
of the neutrino mass.

\section{Direct Leptogenesis Induced by Complex Phases}

We now turn to the second method of generating a CP asymmetry in the
leptonic sector, namely by means of a transient MNS(P) matrix.

Recall first how CP violation comes about in the SM. Here one has the
Yukawa couplings $\frac12 Y_{ij} {\bar\psi}_i (1 + \gamma_5) \psi_j
\phi$, where $\psi$ can be a quark or a lepton field and $\phi$ is the
ordinary Higgs field. Because there are three generations, the Yukawa
couplings $Y_{ij}$ are complex. When $\phi$ acquires a vacuum
expectation value, one develops (complex) mass terms. Their
diagonalization leads to the CKM matrix for quarks, or the MNS(P)
matrix for leptons.

In the present case, we again have couplings of the form $\frac12
Y_{ij} {\bar\psi}_i (1 + \gamma_5) \psi_j \phi_{1,2}$, where the
$\phi_{1,2}$ are the background scalar fields. Here we assume that CP
is conserved in the Lagrangian, so that the $Y_{ij}$ are real.
However, in dynamical CP violation, the initial conditions of the
scalar fields have relatively complex values (e.g.\ see
Eq.(\ref{initc})]. These initial conditions effectively lead to
complex (non-diagonal) Yukawa couplings, which again lead to mass
matrices whose diagonalization includes CP-violating phases. However,
in contrast to the SM, once the fields relax to their minima, the
contributions to the masses are switched off, and the CP-violating
mixing matrix vanishes. In other words, the mixing matrix is {\it
transient}, and no dynamical CP violation remains at late times.

Of particular interest to us is the case where $\psi$ is the
neutrino. Here nonzero neutrino masses are generated, effectively
leading to a MNS(P)-type matrix. This matrix can asymmetrically
produce $\nu_L$ and ${\bar\nu}_R$ even when no lepton-number-violating
interactions are present in the theory.

\vs {\bf Transient MNS(P) Matrix:} The above qualitative description
can be made quantitative. As before, we take the background scalar
fields to be SU(2) doublets (other representations can be
straightforwardly included) and assume only the neutral components
acquire background values. The general form for the Yukawa
interactions is
\be
{\cal L}_y \, = \, \frac12 d_e {\bar e} (1 + \gamma_5) e D^*_b +
\frac12 d_\nu {\bar\nu} (1 + \gamma_5) \nu_R D + h.c.~,
\label{sdtlag}
\ee
where $D$ represents the neutral components of the doublet fields.
Note that $d_e$ and $d_\nu$ are $3\times 3$ Yukawa flavor matrices
which become complex due to initial conditions.

Let us first consider the case of Dirac fermions. The induced complex
Dirac-type masses can then be parametrized as
\be
M_{AB} = a_2 d_{AB}^2 + a_1 e^{i\alpha} d^1_{AB} ~,
\ee 
where $d^i_{AB}$ are the Yukawa couplings for the doublets $\phi_i$,
and $A,B$ are the flavor indices. Since the $d_{AB}^2$ are different
from $d_{AB}^1$, $M_{AB}$ is a generic arbitrary complex $3\times 3$
matrix that needs to be diagonalized using bi-unitary transformations.
Therefore, for a given leptonic doublet, the charged and neutral
fermion mass matrices read as
\be
M_e=U^\dagger_e M_e^d V_e \mbox{ and }M_{\nu} =
U^\dagger_{\nu}M_\nu^d V_{\nu}~,
\label{cmass}
\ee
where the superscript $d$ denotes diagonal elements.  One finds as
usual the MNS(P) matrix
\be
U^\dagger_e U_{\nu}~,
\label{ckm}
\ee
which obviously includes complex CP-violating interactions. 

To see this CP violation, we define the matrices $K_e$ and $K_\nu$
such that
\be
U_e K_e V_e^\dagger = (M_e^d)^2 ~\mbox{ and }(M_\nu^d)^2 =
U_\nu K_\nu^d V_{\nu}^\dagger~.
\label{hermk}
\ee
The condition for CP violation is derived to be \cite{zing}
\be 
C_{ij} = Im(K_{e ij} K_{\nu i j}^*) + Im(K_{e i k} K_{\nu i k}^*) \neq
0 ~;~ i < j~; ~j \neq k ~.
\label{cpcond}
\ee

It is convenient to assume that the matrix describing charged-lepton
couplings ($U_e$) is diagonal, so that the mixing matrix comes purely
from the neutrino sector: $U_{MNS(P)} = U_\nu$. The largest
contribution to CP violation then comes from the case where the
charged lepton is a $\tau$. For a specific leptonic flavor (say
$\alpha$), the condition (\ref{cpcond}) translates to
\be
C_{\tau \alpha} \approx 2 \, m_\tau^2 Im(K_{\nu \tau \alpha}^*) \neq 0 ~.
\label{appcond}
\ee

\begin{figure}
\centerline{\epsfxsize=1.9 in \epsfbox{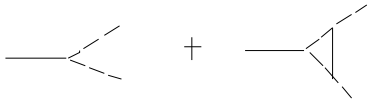}}
\caption{Feynman diagrams of the tree and loop processes considered
for the decay, $\phi_2 \to \bar\nu_{\tau R} \nu_{\tau L}$. The dashed
lines is for fermions and the solid line is for scalar.}
\label{fig1}
\end{figure}

The phase in the neutrino mixing matrix is now responsible for CP
violation in scalar decays. For illustration, let us consider a flavor
specific decay of the type
\be
\phi_2 \to \bar\nu_{\tau L} \nu_{\tau L} ~~,~~~~ \phi_2^\dagger \to
\bar\nu_{\tau R} \nu_{\tau R} ~.
\label{mudecay}
\ee
This decay can be CP-violating because of a phase arising from the
one-loop vertex-correction graph. This process is described in
Fig.\ref{fig1}. The resulting CP-violating asymmetry is
\be
Y_\nu \propto Arg (d_{\tau \tau}) - Arg(d_{e \tau} d_{e \mu} d_{\mu
\tau}^*) = \sin\delta~,
\label{dcpyuk}
\ee
where $\delta$ is the complex (Dirac) phase in $U_\nu$.

Several remarks are in order. First, the asymmetry in (\ref{dcpyuk})
is generated solely due to mixing in the leptonic sector. This is
easily seen by the presence of different flavor-changing Yukawa
couplings in the expression for the asymmetry. Second, we require a
CP-conserving phase in the interference between tree and one-loop
diagrams. This comes about due to an absorptive piece in the loop
calculation. Finally, the asymmetry is prevented from being washed out
due to the same condition as that of (\ref{cond}). As before, the
lepton-number asymmetry is converted to a baryon-number asymmetry
using sphaleron effects.

\vs {\bf Majorana fermions:} We now mention briefly the more
complicated case where the neutrinos are Majorana particles. In
comparison to the Dirac case, the Majorana fermions (due to the
reality of the fields) have two more CP-violating phases (for three
generations) in the mass matrix. These additional phases give further
avenues for CP violation. The important feature that distinguishes
this from the Dirac case is the possibility that $C_{\alpha \alpha}$
[Eq.~(\ref{cpcond})] can take nonzero values \footnote{In fact, this
feature enables one to search for CP violation in neutrino-less double
beta decay experiments \cite{nudbd}.}.

To illustrate this, consider just two generations. In this case, there
is no CP violation for Dirac neutrinos, but there can be for Majorana
neutrinos. The effective Majorana mass is
\be
m_{\alpha \alpha} = m_1 c_1^2 + m_2 s_1^2 e^{2i\gamma}~,
\label{majomass}
\ee
where, $c_1/s_1$ are the cosine/sine of the mixing angle and $\gamma$
is the CP-violating phase. Thus, in this case, one can obtain
CP-violating lepton asymmetries as above in the decays $\phi_2 \to
\bar\nu_L \nu_L$ and $\phi_2^\dagger \to \bar\nu_R \nu_R$. However,
because of their Majorana nature, it is necessary in this case that
the final-state neutrinos have different flavors.

Note that, for Majorana neutrinos, we can replace our scalars $\phi_i$
with those used in the standard thermal leptogenesis scenario
\cite{fuku}; an extension to the nonthermal case \cite{shafi} is
straightforward.

\section{Estimating Baryon asymmetry in a specific example}

\noindent {\bf Relating $\alpha$ to the MNS(P) phase:} Up to this
point, we have concentrated exclusively on showing that the CP
violation in the scalar sector can be successfully transferred to the
fermion sector. However, one might wonder whether this CP violation is
sufficient to obtain the right amount of baryogenesis, i.e.\ a correct
baryon to entropy ratio. In this section we explore this question.

In order to examine this, we need a framework for obtaining CP
violation in the fermion sector. For this, we use the transient MNS(P)
matrix with Dirac neutrinos. In Sec.~4, we showed that CP violation
occurs with two scalar doublets and three generations. However, it is
also possible to obtain a transient MNS(P) matrix with only two
generations, if one considers a left-right model, i.e.\ the gauge
symmetry is extended to $G = SU(2)_L\times SU(2)_R$. This is the
scenario we adopt here.

As discussed earlier, the neutrinos couple to the rolling scalar
doublets. We write the background fields in the form
\be 
\phi_1(t)=c_1(t)~\mbox{ and }~\phi_2(t)=c_2(t)e^{i\a}~.  
\ee 
In the left-right model, these two fields are combined into a single
matrix:
\be
\Phi = \left( \begin{array}{cc} c_2e^{i\a} & 0\\ 0 & c_1
\end{array} \right)~.
\label{srep}
\ee

We focus only on the leptonic sector. The Yukawa interactions for the
lepton doublet $L$ are
\be
{\cal L}_Y = \frac12 y_l \bar L \Phi (1+\gamma_5) L + \frac12 y_\nu
\bar L {\tilde \Phi} (1 + \gamma_5) L~;~\tilde \Phi = \sigma_2 \Phi^*
\sigma_2~.
\label{yukawa}
\ee
The mass matrices for the charged-lepton and neutrino fields are
therefore
\bea
M_l &=& M_l^0 + r M_\nu e^{-i\a}~,\nonumber\\
M_\nu &=& rM_l e^{i\a} + M_\nu^0~,\nonumber\\
r &=& \frac{c_1(t)}{c_2(t)}~.
\label{massmat2}
\eea

In the absence of phases which break CP explicitly, the mass matrices
$M_{l,\nu}^0,M_{l,\nu}$ are real and symmetric (this is an advantage
of choosing the group $G$). As a result, the mass matrices can be
diagonalized by a single mixing matrix $V_{l,\nu}$ (and the left- and
right-handed mixing matrices are related, $U_L = U_R^*$):
\be
V_{l,\nu}M_{l,\nu} V_{l,\nu}^T = M_{l,\nu}^{diag}~.
\label{mixmat1}
\ee
The corresponding MNS(P) matrix is
\be
U = V_l V_\nu^\dagger~.
\label{ckmmat}
\ee

With two generations, apart from an overall phase, the MNS(P) matrix
can be parametrized as
\be
U = \left( \begin{array}{cc}
e^{i\delta_1} \cos\theta & e^{i\delta_2} \sin\theta \\
-e^{-i\delta_2} \sin\theta & e^{-i\delta_1} \cos\theta
\end{array} \right)~.
\label{param1}
\ee
Note that, although the above complex mass matrix is written in terms
of the two phases $\delta_{1,2}$, only the difference $\delta_1 -
\delta_2$ is physical.

For calculational ease we assume that $c_1(t) \ll c_2(t)$. In this
case, $r \ll 1$ for any given time, and thus $r(t)$ can be treated as a
perturbative parameter. In this limit, the relation between the phase
$\alpha$ and the complex phase difference $\delta (\equiv \delta_2 -
\delta_1)$ is given approximately by \cite{chang}
\be
\delta \equiv \delta_1 - \delta_2 \simeq \frac{m_\tau}{m_\nu} r
\sin\alpha~.
\label{ckmph}
\ee
In (\ref{ckmph}) $m_\nu$ is one of the mass eigenvalues of the
neutrino sector which we take to be considerably smaller than the tau
lepton mass, $m_\tau$. Note that the MNS(P) phase is time dependent as
long as $r$ is time dependent.

In the following we estimate this time dependence for $r$. This will
allow us to determine whether the proper baryon to entropy ratio is
obtained.

\vs {\bf Cosmological Evolution and the Baryon \\ \null
~~~~~Asymmetry:}

As in all models of leptogenesis, in our scenario the net baryon
number is generated via sphalerons from an asymmetry in the
left-handed lepton number. As long as sphaleron transitions are
thermodynamically allowed, i.e.\ at energy scales higher than the
electroweak scale, they will equilibrate the baryon and left-handed
lepton numbers \cite{KRS}. In our context, this means that a baryon
asymmetry of the same order of magnitude as the left-handed lepton
asymmetry produced by the processes described in the previous
subsection will be generated. In turn, the left-handed lepton
asymmetry is generated by the motion of the scalar field $\phi$. We
will estimate the net baryon to entropy ratio by the ratio of
left-handed leptons over photons produced during the time interval
when the rolling of the scalar field is most important in the sense
discussed below. Note that, once produced, the baryon to entropy ratio
does not change in time as long as there are no other phase
transitions which generate entropy after the period of
leptogenesis. Thus, in the following we will first determine the time
interval when most of the left-handed leptons are produced, and then
estimate the resulting baryon to entropy ratio following the methods
of \cite{bala2,AD}.
  
The equation of motion for any rolling complex scalar field which
is homogeneous in space is
\be
\ddot{\phi} + 3H\dot{\phi} \, = \, -\frac{\d V(\phi_i)}{\d \phi} \, ,
\label{eom1}
\ee
where $H$ is the Hubble parameter. The dynamics of $\phi$ depend
on whether $H(t)$ is larger or smaller than $m_{\phi}$, the mass
of the field $\phi$ which is given by the square root of the second
derivative of the potential $V(\phi)$.

At early times, while
\be \label{constraint1}
H(t) \, > \, m_{\phi} \, ,
\ee
the evolution of $\phi$ is over-damped and follows the approximate
equation
\be \label{eom2}
3H\dot{\phi} \, = \, -\frac{\d V(\phi_i)}{\d \phi} \, 
= \, -m_{\phi}^2 \phi \, .
\ee
In the early radiation-dominated universe
\be \label{Hubble}
H(t) \, = \, {1 \over {2t}} \, ,
\ee
and thus the solution of (\ref{eom2}) becomes
\be
\phi(t) \, = \, \phi(t_0) e^{-{1 \over 3} m_{\phi}^2 (t^2 - t_0^2)} \, ,
\ee
where $t_0$ is the initial time. Making use of (\ref{Hubble}) and
(\ref{constraint1}), we see that the motion of $\phi$ in this phase
is negligible. Hence, no leptogenesis takes place during this phase.

The inequality (\ref{constraint1}) is saturated at a time we denote
as $t_e$. As soon as
\be \label{constraint2}
H(t) \, \leq \, m_{\phi} \, ,
\ee
the scalar field $\phi$ starts rolling as described by (\ref{eom1})
and eventually performs damped oscillations about its ground state.
The sign of the induced lepton asymmetry depends on the sign of
${\dot \phi}$. Thus, if $\phi$ were oscillating with constant amplitude,
no net lepton number would be generated. The net lepton asymmetry
is determined by the overall decrease in the amplitude of $\phi$.
Most of this decrease happens during the time interval before $\phi$
crosses zero for the first time, which occurs at a time we denote
by $t_*$. The time interval between $t_e$ and $t_*$ is less than one
Hubble expansion time $H(t_e)^{-1}$.

For the baryogenesis channel considered in the above subsection, the
net baryon to entropy ratio is given by \cite{AD,bala2}
\be \label{final1}
{{n_B} \over s} \, \sim \, {{y_{\nu}^2 Y_{\nu} (\phi {\dot \phi})(t_e)}
\over {8 \pi n_{\gamma}(t_e)}} \, ,
\ee
where $y_{\nu}$ is the typical value of a neutrino Yukawa coupling
constant from (\ref{yukawa}), $Y_{\nu}$ is the CP asymmetry per decay
from (\ref{dcpyuk}), $n_{\gamma}$ is the number density of photons,
$s$ denotes the entropy density, and $n_B$ the net baryon number
density. A way to understand the above equation is as follows:
$m_{\phi} Y_{\nu}$ is the CP asymmetry in the decay rate of $\phi$
quanta, $(\phi {\dot \phi})(t_e)$ is the rate of change in the number
of $\phi$ quanta at time $t_e$, the time when most of the net baryon
number density is generated, and $m_{phi}^{-1}$ gives the time interval
during which leptogenesis is effective and must be multiplied with the
rate of generation of the baryon asymmetry to obtain the final
baryon to entropy ratio.

In turn, the asymmetry factor $Y_{\nu}$ in the above equation (\ref{final1})
is given by combining (\ref{dcpyuk}) and (\ref{ckmph}) and inserting the
time evolution of the function $r(t)$. However, given the gauge symmetry
of our model, the masses of both scalar fields $\phi_1$ and $\phi_2$ are
the same and they will thus begin rolling at the same time with vanishing
velocity. Hence, $r(t)$ will be constant in time. Since we are also taking
the phase $\alpha$ to be constant, the phase $\delta$ and hence $Y_{\nu}$
are independent of time.

To evaluate the order of magnitude of the result (\ref{final1}), we must
estimate the value of ${\dot \phi}(t_e)$. From the equation of motion (\ref{eom1})
it follows that
\be
{\dot \phi}(t_e) \, \sim \, m_{\phi} \phi(t_e)
\ee
and hence
\be \label{final2}
{{n_B} \over s} \, \sim \, {{y_{\nu}^2 Y_{\nu} m_{\phi} \phi^2(t_e)}
\over {8 \pi n_{\gamma}(t_e)}} \, .
\ee
Since the time $t_e$ is given by the saturation of (\ref{constraint1}), the
photon number density is given by
\be
n_{\gamma}(t_e) \, \sim \, m_{\phi}^{3/2} m_{pl}^{3/2} \, ,
\ee
where $m_{pl}$ is the Planck mass, and we have used the Friedmann equations to relate
the temperature of the radiation bath (whose cube yields $n_{\gamma}$) to the
Hubble expansion rate at time $t_e$ which in turn is equal to $m_{\phi}$.
Thus, (\ref{final2}) becomes
\be \label{final3}
{{n_B} \over s} \, \sim \, y_{\nu}^2 Y_{\nu} {{\phi^2(t_e)} \over 
{m_{\phi}^{1/2} m_{pl}^{3/2}}} \, .
\ee

We now add an additional constraint from cosmology: we demand that the
energy density in $\phi$ at time $t_e$ be sub-dominant, i.e.
\be \label{bound}
m_{\phi}^2 \phi^2(t_e) \, \ll \, T^4(t_e) \, \sim \, H^2(t_e) m_{pl}^2 \, ,
\ee
where $T(t)$ denotes the temperature of radiation, and where in the second step
we have again used the Friedmann equations. Recalling that $H(t_e) = m_{\phi}$,
the inequality (\ref{bound}) yields an upper bound on $\phi(t_e)$:
\be \label{bound2}
\phi(t_e) \, \ll \, m_{pl} \, ,
\ee
a bound which is quite natural from the point of view of particle physics (we 
cannot trust the physics we used for field values larger than $m_{pl}$).
Inserting the bound (\ref{bound2}) into (\ref{final3}), we obtain a bound on
the strength of our baryogenesis scenario of the form
\be \label{final4}
{{n_B} \over s} \, \ll \, 
y_{\nu}^2 Y_{\nu} \bigl({{m_{pl}} \over {m_{\phi}}}\bigr)^{1/2} \, .
\ee
This result demonstrates that our dynamical CP violation scenario can lead,
even in the case of small coupling constants, to a large net baryon to
entropy ratio.

\section{Other Channels for Lepto- and Baryogenesis}

There are other ways of transferring a CP asymmetry in the scalar
sector to the lepton sector. In this section, we briefly describe some
alternative mechanisms.

\begin{figure}
\centerline{\epsfxsize=1.9 in \epsfbox{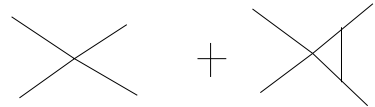}}
\caption{Feynman diagrams of the tree and loop process for the reaction
$\phi_1 \phi_2 \to 2 \phi_2$.}
\label{fig2}
\end{figure}

\vs {\bf Asymmetry due to scattering:} In Sec.~3, we considered the
possibility that a CP-violating asymmetry in the scalar sector can be
generated in the decay $\phi_1 \to \phi_2 \phi_2$. This is then
converted to a CP asymmetry in the lepton sector via the decay of
$\phi_2$ scalars into neutrinos. However, a CP asymmetry in the scalar
sector can also be created by scattering processes such as $\phi_1
\phi_2 \to 2 \phi_2$. This process is generated by the quartic
interaction $V_4$ in (\ref{pot2}) and has an one-loop vertex
correction due to the cubic interaction $V_3$ generated in
(\ref{cubic2}). Once again, the asymmetry comes about because the
coupling of $\phi_2$ to $\phi_1$ is complex, but $\phi_2^3$
interactions are real. This leads to a phase difference between the
tree diagram and the one-loop vertex correction graph, and contributes
to the asymmetry. The relevant Feynman graph for this process is shown
in Fig.\ref{fig2}.

The asymmetry which is generated between the cross-sections for
$\phi_2$ and $\phi_2^*$ production (via the CP-conjugate process) is
\be
\sigma - \bar\sigma \sim (\mbox{loop factor})\times \sigma \sin\alpha~;~
\sigma = \frac{1}{32\pi} \frac{g^2}{s}~.
\label{phi2asym}
\ee
In (\ref{phi2asym}) $s$ is the center of mass energy for the
scattering process and $g$ is the relevant quartic coupling in
(\ref{pot2}). Following (\ref{phi2asym}), the asymmetry in the rate of
$\phi_2$ production is
\be
\Delta \Gamma_{\phi_2} = (\sigma - \bar\sigma) \cdot N_{\phi_2}
= \frac{\sin\alpha}{32\pi} \frac{g^2}{s}\cdot N_{\phi_2}~,
\label{rateasym}
\ee
where $ N_{\phi_2}$ is the number density of the scattering of
$\phi_2$ particles \footnote{For simplicity, we assume equal number of
$\phi_1$ and $\phi_2$ fields.}.

Once again, in this case, the lepton asymmetry is generated by
allowing the scalar $\phi_2$ to decay via the process $\phi_2 \to
\nu_L {\bar\nu}_L$. The lepton number asymmetry generated in this way
is proportional to the Yukawa coupling [(\ref{dasym})]:
\be
Y_\nu \propto d_{\nu 2}^2 \Delta \Gamma_{\phi_2}= 
\frac{\sin\alpha}{32\pi} \frac{(gd_{\nu 2})^2}{s}\cdot N_{\phi_2}~.
\label{dasyms}
\ee
The interesting aspect of (\ref{dasyms}) is that the asymmetry could be large
for large values of $ N_{\phi_2}$, even if the Yukawa coupling is small. 
Simultaneously, the wash-out factor is not large because the inverse scattering
may be kinematically suppressed if the initial masses are large compared to the
final state masses. In our case, we have $m_1 > m_2$ and we expect
$ m_i \gg a_2 d_{\nu 2}$ due to small Yukawa couplings.

\begin{figure}
\centerline{\epsfxsize=1.9 in \epsfbox{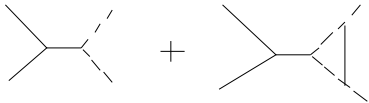}}
\caption{Feynman diagrams of the tree and loop process for the reaction
$\phi_1 \phi_2 \to \nu_L {\bar\nu}_R$.}
\label{fig3}
\end{figure}

One can also produce a lepton-number asymmetry directly via the pair
production of Dirac fermions in the scattering process $\phi_1 \phi_2
\to \nu_L {\bar\nu}_L$. The relevant diagrams are shown in
Fig.\ref{fig3}.  Taking the final state to be $\nu_{\tau L}
\bar\nu_{\tau L}$, we obtain an asymmetry which is approximately given
by
\be
Y_\nu \propto g|d_{\tau\tau}d_{e \tau} d_{e \mu} d_{\mu \tau}|
\frac{\sin\delta}{16\pi s} \cdot N_{\phi_2}~.
\label{dasymf}
\ee
The asymmetry in (\ref{dasymf}) is smaller that that generated in
(\ref{dasyms}) due to a larger Yukawa suppression.

Finally, a word about Majorana neutrinos. The above asymmetry which we
have generated is suppressed if we use Majorana neutrinos. This
follows due to spin statistics, where we notice that our scalars have
to generate final states in the P-wave mode and hence the scenario is
disfavored if it involves Majorana neutrinos.

\vs {\bf Exotic fermions:} Finally, it is also possible to construct
scenarios involving the use of exotic fermions to transfer the CP
violation to the fermion sector. Although we will not discuss such
models in any detail, there are many possibilities. For example, the
scalar fields $\phi_i$ can couple to such (neutral or charged) exotic
fermion fields, so that the CP asymmetry in the $\phi_i$ is then
transferred to the exotic fermion sector. The subsequent decays of
exotic fermions into ordinary fermions creates the necessary
lepton-number asymmetry.

\section{Discussion and Conclusions}

In this paper we have elaborated on the scenario of dynamical
cosmological CP violation we proposed in \cite{bala1} (see also
\cite{bala2}). In this scenario, the CP violating-phases are due to
cosmological initial conditions for some new scalar fields $\phi_i$
(which could be some of the moduli fields emerging from new physics at
very high energies). Here, we have focused on how to transfer the CP
asymmetry from the scalar sector to the leptonic sector, and on the
connection with baryogenesis.

We mention several ways to achieve the transfer of the CP asymmetry to
the leptonic sector. One intriguing possibility is that Yukawa
couplings between the scalar field $\phi_i$ and the standard model
leptons could generate a transient (in time) MNS(P) matrix which
generates a CP asymmetry in the leptonic sector, yielding an
implementation of leptogenesis. In this model, we estimate the
resulting net baryon to entropy ratio. As expected, the result depends
on the initial values of the scalar fields. We find, however, that it
is easy to generate a sufficiently large baryon asymmetry to explain
the data, even if the coupling constants and the value of the CP
violating phase in the scalar sector are small.

Note that our scenario does not assume any new sources of CP and
baryon-number violation in the Lagrangian. The CP violation comes from
the scalar field initial conditions, and the baryon asymmetry is
generated via sphalerons from an asymmetry in the left-handed leptons,
an asymmetry which is compensated by an asymmetry in the right-handed
leptons. A prediction of this scenario is that there should be an
asymmetry in the number of right-handed neutrinos commensurate with
the observed net baryon to entropy ratio.

{\bf Acknowledgements}:

This work is supported by NSERC (Canada) and by the Fonds de Recherche
sur la Nature et les Technologies du Qu\'ebec. 
RB is also supported in part by the US Department of Energy under Contract
DE-FG02-91ER40688, TASK~A.  


\end{document}